\documentclass[12pt]{emulateapj}

\usepackage{amsmath,amssymb,natbib,graphicx}
\usepackage[usenames]{color}
\usepackage[colorlinks,urlcolor=blue,citecolor=black,linkcolor=blue]{hyperref} 

 \newcommand{\nuvr}{{$(\rm{NUV} - r^+)$ }}

\shorttitle{Transformers}
\shortauthors{George et al.}
 
\begin{document}
  

 \title{Galaxies in X-ray Groups. III. Satellite Color and Morphology Transformations}



\author{
Matthew R. George\altaffilmark{1,2},
Chung-Pei Ma\altaffilmark{1},
Kevin Bundy\altaffilmark{3}, 
Alexie Leauthaud\altaffilmark{3}, 
Jeremy Tinker\altaffilmark{4},
Risa H. Wechsler\altaffilmark{5,6},
Alexis Finoguenov\altaffilmark{7},
Benedetta Vulcani\altaffilmark{3}
}

\altaffiltext{1}{Department of Astronomy, University of California,
  Berkeley, CA 94720, USA}
\altaffiltext{2}{Lawrence Berkeley National Laboratory, 1 Cyclotron
  Road, Berkeley, CA 94720, USA}
\altaffiltext{3}{Kavli Institute for the Physics and Mathematics of
  the Universe (Kavli IPMU, WPI), Todai Institutes for Advanced Study,
University of Tokyo, Kashiwa 277-8583, Japan}
\altaffiltext{4}{Center for Cosmology and Particle Physics, Department
of Physics, New York University, 4 Washington Place, New York, NY
10003, USA}
\altaffiltext{5}{Kavli Institute for Particle Astrophysics and
  Cosmology, SLAC National Accelerator Laboratory, 2575 Sand Hill
  Rd., Menlo Park, CA 94025, USA}
\altaffiltext{6}{Physics Department, Stanford University, Stanford, CA
94305, USA}
\altaffiltext{7}{Department of Physics, University of Helsinki, Gustaf
H\"{a}llstr\"{o}min katu 2a, FI-00014 Helsinki, Finland}

\email{mgeorge@astro.berkeley.edu}


  
\begin{abstract}
  While the star formation rates and morphologies of galaxies have
  long been known to correlate with their local environment, the
  process by which these correlations are generated is not well
  understood. Galaxy groups are thought to play an important role in
  shaping the physical properties of galaxies before entering massive
  clusters at low redshift, and transformations of satellite galaxies
  likely dominate the buildup of local environmental correlations. To
  illuminate the physical processes that shape galaxy evolution in
  dense environments, we study a sample of 116 X-ray selected galaxy
  groups at $z=0.2-1$ with halo masses of $10^{13}-10^{14}~{\rm
    M_{\odot}}$ and centroids determined with weak lensing. We analyze
  morphologies based on \textit{HST} imaging and colors determined
  from 31 photometric bands for a stellar mass-limited population of
  923 satellite galaxies and a comparison sample of 16644 field
  galaxies. Controlling for variations in stellar mass across
  environments, we find significant trends in the colors and
  morphologies of satellite galaxies with group-centric distance and
  across cosmic time. Specifically at low stellar mass
  ($\log(M_{\star}/M_{\odot}) = 9.8-10.3$), the fraction of
  disk-dominated star-forming galaxies declines from $>50\%$
  among field galaxies to $<20\%$ among satellites near the centers of
  groups. This decline is accompanied by a rise in
  quenched galaxies with intermediate bulge+disk morphologies, and
  only a weak increase in red bulge-dominated systems. These results
  show that both color and morphology are influenced by a galaxy's
  location within a group halo. We suggest that strangulation and disk
  fading alone are insufficient to explain the observed morphological
  dependence on environment, and that galaxy mergers or close tidal
  encounters must play a role in building up the population of
  quenched galaxies with bulges seen in dense environments at low
  redshift.
\end{abstract}
 



\section{Introduction}

Galaxy properties are correlated with their environment. Dense regions
host galaxies with greater masses, lower star-formation rates, and
morphologies that are more bulge-dominated than in low density
regions.  Since the early work of \citet{Dressler1980}, studies of
environmental correlations have expanded to a variety of indicators of
star formation, morphology, and environment. See \citet{Blanton2009}
for a recent review of these correlations in the local universe.

Numerous processes may be culpable for the dependence of galaxy
properties on environment (see \citealt{Boselli2006} for a
review). Galaxy interactions through mergers or tides can disrupt
stellar kinematics and remove gas. These interactions can also apply
torques that drive gas inward, perhaps feeding star formation or a
central black hole. Halos can play a role through tidal forces
and dynamical friction. Pressure from hot dense gas inside halos
may strip gas from infalling galaxies, and shock heating
in massive halos can prevent accretion of cold gas that would feed
star formation.
 
Despite extensive observational and theoretical work, the dominant
physical mechanisms responsible for the environmental correlations
remain unclear.  An important clue is that the scale on which these
correlations appear is similar to the size of the dark matter halos
hosting galaxies \citep[e.g.,][]{Kauffmann2004,
  Blanton2007}. Relatedly, satellite galaxies have a primary role in
the buildup of the red sequence in dense environments
\citep[e.g.,][]{Weinmann2006, vandenBosch2008, Wetzel2012a, Peng2012}.  

Since many galaxy characteristics are interrelated, it is important to
study the correlations for each property independently while fixing
other variables. When constraining environmental effects, one ideally
controls for differences in redshift, stellar mass, halo mass, and
location within a group. The colors and morphologies of galaxies are
also correlated, so it is advantageous to split galaxies
simultaneously by color and morphological classes to distinguish
between processes that affect star formation rates and structural
properties differently. Several studies have suggested that
environmental processes affect star formation more significantly than
morphology \citep[e.g.,][]{Kauffmann2004, Blanton2005, Christlein2005,
  Weinmann2009, Kovac2010},
suggesting that the gas that feeds star formation is depleted or
removed without significantly altering galactic structure or stellar
kinematics.

While many of the observational studies have focused on large surveys
at low redshift or on massive clusters at higher redshifts, less is
known about the more common group-scale environments and their
evolution over time. In this paper, we study the colors and
morphologies of galaxies in groups spanning the redshift range
$z=0.2-1$. We focus in particular on the properties of satellite
galaxies as a function of stellar mass, group-centric distance, and
redshift. This sample of groups has been identified from the COSMOS
field \citep{Scoville2007} based on their extended X-ray emission and
have masses determined with weak lensing in the range $10^{13}-10^{14}
M_{\odot}$ \citep{Leauthaud2010}. Galaxy membership has been assigned
using precise photometric redshifts with a Bayesian procedure
tested extensively with mock catalogs and a spectroscopic subsample
\citep{George2011}. Furthermore, by finding galaxies that maximize the
lensing signal on small scales we have identified central galaxies
that accurately trace the center of mass of their halos
\citep{George2012}. Additional studies of these
  groups with magnification lensing
\citep{Ford2012, Schmidt2012} and clustering \citep{Allevato2012} have
confirmed the halo mass estimates. These ingredients make for a robust
characterization of group environments with which to study satellite
properties and their evolution over time.

We describe the data used in this analysis in Section~\ref{s:data}
along with mock catalogs used to estimate the purity of the satellite
population analyzed. Section~\ref{s:results} presents the results
including our primary finding of a declining fraction of blue late
disk galaxies and a rise in red early types toward group centers and
over cosmic time. In Section~\ref{s:discussion}, we discuss the
implications of these findings for the physical mechanisms that shape
galaxy evolution, and put this work in the context of previous studies
and possibilities for the future.

\section{Data and Mocks}
\label{s:data}

The COSMOS field \citep{Scoville2007} has been the subject of a broad
array of multi-wavelength imaging and spectroscopy campaigns. In this
study we make use of several pieces of this data: galaxy colors, photometric redshifts, and stellar
masses derived from over thirty bands of ultraviolet, optical, and
infrared data \citep{Ilbert2009, Ilbert2010, Bundy2010}; galaxy
morphologies determined by \citet{Scarlata2007} from high-resolution imaging taken by the
Advanced Camera for Surveys (ACS) aboard the \textit{Hubble Space
  Telescope} \citep{Koekemoer2007}; and a catalog of massive galaxy
groups selected based on their X-ray emission seen with
\textit{XMM-Newton} and \textit{Chandra} and for which halo mass
profiles have been characterized using weak lensing with
\textit{Hubble} imaging \citep{Leauthaud2010, George2011,
  George2012}. Most of these data have been compiled in the group
member catalog published in \citet{George2011} and we refer the reader
to that paper and references therein for further details including the
flux and mass limits for galaxies and groups. Here we briefly explain
the stellar masses, colors, and morphologies used in this paper, as
well as the identification of group centers and the mock catalogs used
for validating group membership assignment and estimating the effects
of contamination from field galaxies.

\subsection{Group Membership and Mocks}
\label{s:groupdata}

As a measure of galaxy environment, we use a catalog of groups
described in \citet{George2011}. These groups have been identified
through a wavelet detection of extended X-ray emission following
\citet{Finoguenov2010} and the sample is an expanded version of an
earlier COSMOS catalog \citep{Finoguenov2007} with deeper X-ray data.
A Bayesian membership algorithm is used to associate galaxies with
groups based on precise photometric redshifts and proximity to the
X-ray position, with a prior accounting for the relative field
population as a function of magnitude and redshift. Following
\citet{George2011}, we select galaxies with membership probability
$P_{\rm mem}>0.5$ among the clean sample of groups, avoiding masked
regions, groups with fewer than four members, overlapping groups, and
possible projections (\textsc{flag\_include}=1 in the catalog). Halo
masses for these groups have been measured with stacked weak lensing
to be in the range $M_{\rm 200c} \approx 10^{13} - 10^{14} M_{\odot}$
and are correlated with X-ray luminosity \citep{Leauthaud2010,
  George2012}. Over this limited range in halo masses, we have not
detected significant trends in galaxy properties with X-ray
luminosity, but to avoid the influence of outliers we eliminate a
small number of groups with halo masses estimated from
  their X-ray luminosities following \citet{Leauthaud2010} that are
above or below the $10^{13}-10^{14} M_{\odot}$ range.

We split galaxies into central, satellite, and field populations
across several stellar mass and redshift bins to isolate dependences
among different properties. For each group, the central galaxy is
defined as the most massive group member within a projected distance
of the X-ray position equal to the scale radius of a
Navarro-Frenk-White \citep{Navarro1996} profile. \citet{George2012}
showed with weak lensing tests that this definition traces the halo
center to within roughly $75~\rm{kpc}$, though in roughly $30\%$ of
groups the center is still ambiguous. In this paper we do not
specifically study the central galaxies in these X-ray selected halos
because their abundance limits statistical constraints on the
properties of this population. The remaining group members are called
satellites, and are the primary focus of this work. Field galaxies are
taken to be those not assigned to any extended X-ray source and are
thus expected to reside in halos less massive than the X-ray detection
limit ($P_{\rm mem}=0$ from \citealt{George2011}; see Figure 1 of that
paper for the X-ray flux limit as a function of redshift). Because we do
not detect faint X-ray groups at high redshift, the halo mass range
occupied by the field population can evolve due to this selection
effect. However, at low redshift only $\sim10\%$ of galaxies in our
sample live in X-ray groups, so we expect the evolving halo mass limit to have
only a small effect on the statistics of the field
population. Table~\ref{tab:census} gives the size of each of these
samples used in our analysis.

\begin{deluxetable}{crr}
\tablecaption{Environmental Census}
\tablehead{ \colhead{} & \multicolumn{2}{c}{Stellar Mass $[\log(M_{\star}/M_{\odot})]$} \\
\colhead{Type} & \colhead{$[9.8,10.3)$} & \colhead{$[10.3,10.8)$}}
\startdata
\sidehead{$z=0.2-0.5; N_{\rm groups}=47$}
Satellites & 237 & 218 \\ 
Field & 2455 & 1993 \\ 
\sidehead{$z=0.5-0.8; N_{\rm groups}=40$}
Satellites & 122 & 189 \\ 
Field & 4497 & 3529 \\ 
\sidehead{$z=0.8-1.0; N_{\rm groups}=29$}
Satellites & ... & 157 \\ 
Field & ... & 4170
\enddata
\tablecomments{No contamination corrections have been
  applied. Ellipses denote a bin below our stellar mass
  completeness limit. \label{tab:census}}
\end{deluxetable}

Extensive tests of the accuracy of the photometric redshifts and
membership assignment algorithm have been carried out using mock
catalogs from simulations as well as a subsample of galaxies with
spectroscopic redshifts \citep{George2011}. The overall purity and
completeness of the member selection within $0.5 R_{\rm 200c}$ is $84\%$
and $92\%$, respectively, down to our
flux limit of F814W~$=24.2$. The
accuracy of member selection depends most significantly on distance
from the group center and on the flux of a galaxy. The first effect is
due to the decreasing density of true members with group-centric
distance, which implies that selecting galaxies uniformly with radius
(out to $R_{\rm 200c}$) will result in higher contamination from field
galaxies in the outskirts. The second effect from galaxy flux arises
from the decreasing precision of photometric redshifts for fainter
galaxies, ranging from $\sigma_z \lesssim 0.01$ at F814W~$< 22.5$ to
$\sigma_z=0.03$ at F814W~$= 24$. It is notable that with the many
filters used in constructing these photometric redshifts, precision
does not depend significantly on spectral type, so red and blue galaxies of a
given magnitude are selected with similar completeness.

In our analyses of member populations, we correct for contamination
from field galaxies using the mock catalogs described in
\citet{George2011}. A halo catalog is constructed from a LasDamas
simulation (C. K. McBride et al., in preparation)\footnote{Details
  regarding the LasDamas ``Consuelo'' simulation can be found at
  \url{http://lss.phy.vanderbilt.edu/lasdamas/simulations.html}} and
populated with group members following the halo occupation model of
\citet{Leauthaud2012}. Mock galaxies are matched to real COSMOS
galaxies in narrow bins of stellar mass and redshift to assign
magnitudes in the F814W band, which in turn are used to assign mock photometric redshift
errors. We estimate the purity of the member selection algorithm in
bins of group-centric distance and magnitude, and assume that the
colors and morphologies of contaminating field galaxies are
representative of the field population at that
magnitude. Contamination corrections are applied to the satellite
populations only, since the field population is much larger. These
corrections only account for contamination of the member list and not
for incompleteness; the latter effect is smaller, does not
significantly depend on group-centric distance, and should be fairly
uniform across galaxy types since the precision of photometric
redshifts is effectively achromatic \citep[see Figures 2 and 5 of][]{George2011}.

\subsection{Stellar Masses, Colors, and Morphologies}
\label{s:otherdata}

Stellar masses and spectral classes have been determined by fitting
stellar population synthesis models to the spectral energy
distributions of galaxies, varying the age, amount of dust extinction,
and metallicity in the models. We use the stellar masses from
\citet{Bundy2010} and color classifications from \citet{Ilbert2010},
both of which are based on \citet{Bruzual2003} models fit with a
\citet{Chabrier2003} initial mass function but take
somewhat different approaches to fitting the data. Using the
unextincted rest-frame color \nuvr of the best-fitting template for
each galaxy, we split them into two color classes:
\begin{itemize}
\item{\textit{red}: \nuvr $>3.5$}
\item{\textit{blue}: \nuvr $<3.5$.}
\end{itemize}
\citet{Ilbert2010} showed that this color cut corresponds with a
specific star formation rate of roughly $10^{-11} \rm{yr^{-1}}$.
It is worth emphasizing that these colors are corrected for dust
extinction, so that the red sample selects a population of quiescent
galaxies similar to those identified using two color cuts based on
near-ultraviolet, optical, and near-infrared measurements, minimizing
contamination from dusty star-forming galaxies
\citep[e.g.,][]{Bundy2010}.

Galaxies are classified morphologically based on automated
measurements of several structural parameters from ACS imaging
in the F814W band. We use the Zurich Estimator of Structural Types
\citep[ZEST;][]{Scarlata2007} catalog, which is derived from
measurements of the asymmetry, concentration, Gini coefficient, second
moment of the brightest pixels, and ellipticity. \citet{Scarlata2007}
performed a principal component analysis of these measured quantities
to identify the most important structural parameters, and divided
galaxies into classes (early type, disk, and irregular) based on their
location in this principal component space. A further subdivision of
the disk category was derived by correlating measurements of the
S\'{e}rsic index for a bright sample with $I_{AB}<22.5$ with location in the principal
component space. This parametrization provides a rough indication of
the bulge-to-disk ratio, with four classes ranging from
bulge-dominated to bulgeless disks. For statistical purposes, we
consolidate some of these classifications into three morphological
categories: 
\begin{itemize}
\item{early type \textit{spheroidals} and bulge-dominated
disks including even relatively inclined S0 galaxies (ZEST types 1 and
2.0)}
\item{intermediate \textit{bulge+disk} galaxies (ZEST type 2.1)}
\item{\textit{late disks} with little or no bulge component (ZEST types 2.2 and
2.3).}
\end{itemize}
Irregulars and unclassified galaxies make up a small fraction of
the sample, typically not more than a few percent in any stellar mass
bin. We note that morphological K-corrections, which have not been
applied, should not have a large effect on our sample because the F814W band probes rest-frame
optical wavelengths over our entire redshift range
\citep[e.g.,][]{Lotz2004, Cassata2005, Bundy2010}.

\section{Results}
\label{s:results}

\begin{figure*}[htb]
\epsscale{1.2}
\plotone{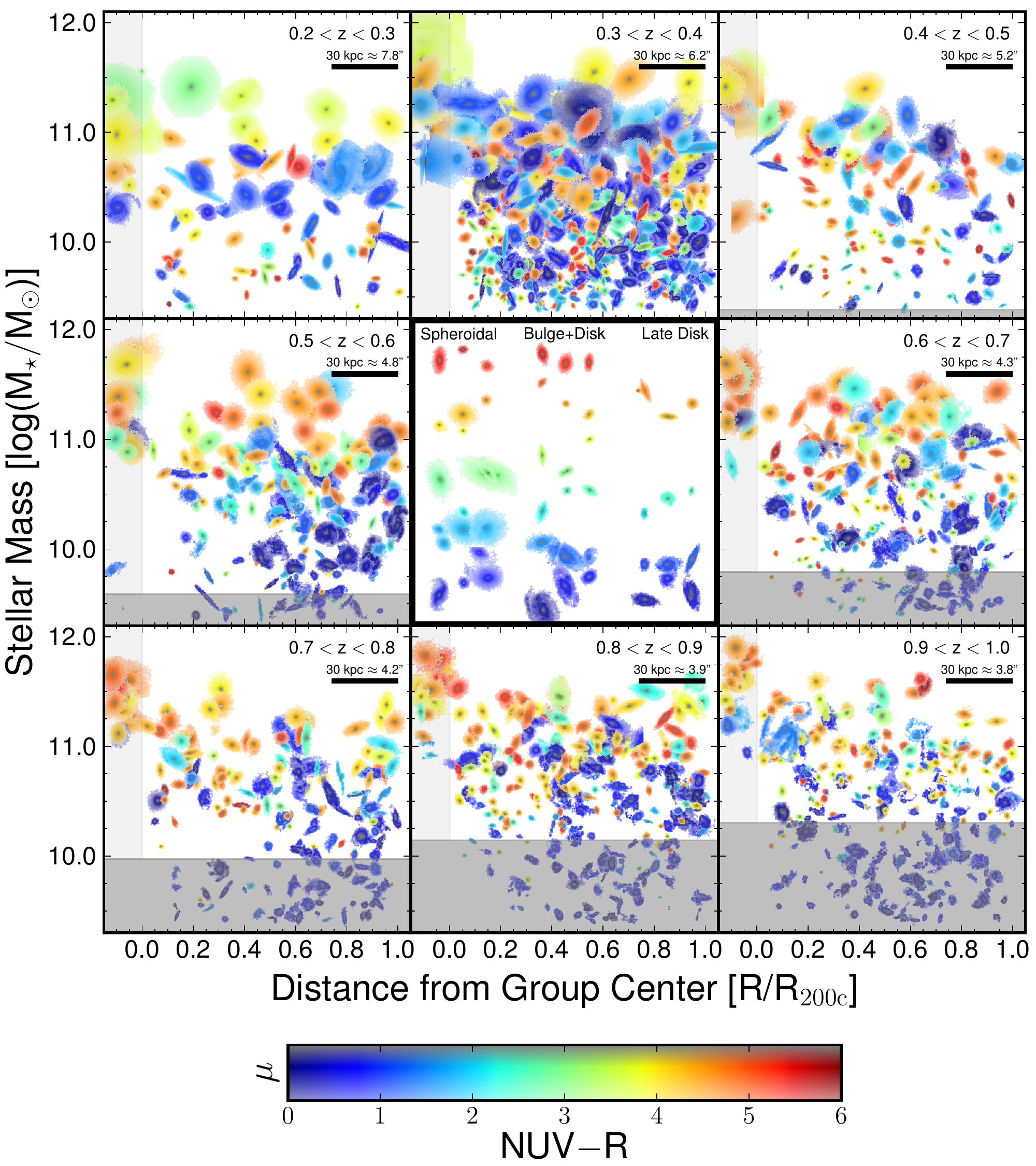}
\caption{Group members as a function of stellar mass and distance from
  the group center. Colors for each galaxy represent the average
  unextincted rest-frame template \nuvr color, with shading
  proportional to the logarithmic surface brightness $\mu$. The gray
  band at the bottom of the high-redshift panels shows the stellar
  mass completeness limit for a passive population calculated with our
  flux limits F814W $ = 24.2$ and $K_s=24$ following the approach of
  \citet{Bundy2010} \citep[see also Figure 1
  of][]{George2011}. Central galaxies in these groups are shown in the
  light gray band on the left side of each of the outer panels. The
  middle frame shows morphological classifications for a random sample
  of galaxies chosen to span the range of colors observed; objects in
  this panel are sorted vertically by color and horizontal offsets
  within each classification are arbitrary.}
\label{fig:candy}
\end{figure*}
 
We begin our analysis with a visualization of a few properties of the
galaxies in our sample. Figure~\ref{fig:candy} shows the distributions
of stellar masses and group-centric distances in different redshift
bins for all galaxies selected as group members. Each point represents
one object and is displayed using the ACS image of the galaxy \citep{Koekemoer2007}, colored
according to the average unextincted rest-frame template \nuvr
color. Centrals are positioned on the left of each panel with small
horizontal offsets, and satellites are plotted at their distance from
the corresponding central. Each image is basically the set of adjacent
pixels with flux above a noise threshold shown with a logarithmic
surface brightness scale that has its maximum set to the peak value
for each galaxy.  We note that the apparent size of a galaxy on the
plot is quite sensitive to its flux since images of brighter galaxies
have more pixels above the noise threshold, and this is not a perfect
indicator of the physical effective radius of a galaxy. A small
fraction of objects have deblending issues or edge effects from the
cutout image size, but we note that the images are processed
independently for visualization and analysis purposes.
 
Several trends are evident when visualizing galaxies in this
manner. Stellar mass is a strong determinant of galaxy properties;
massive galaxies are more likely to be large, red, and
spheroidal. Physical properties also depend on the location within a
group. Central galaxies are massive (by definition), but also
typically red and elliptical. Blue centrals tend to be less massive
than red centrals at high redshifts even within the fairly narrow
range of halo masses studied here, as discovered by
\citet{Tinker2012}. Satellites closer to group centers are more likely
to be red and show fewer spiral features than in the outskirts,
particularly at low stellar masses. There is a relative dearth of low
mass satellites near group centers; this could be a hint of satellite
depletion due to mergers, but challenges in measuring photometry of
faint objects near massive extended central galaxies could be a
contributing factor so further investigation is needed. Similar
evidence of mass segregation and the influence of mass and environment
on star formation has been seen in groups and clusters out to $z\sim
1$ \citep{Muzzin2012, Presotto2012}. Many of these trends can be seen
across the entire redshift range studied here, indicating that both
stellar mass and halo environment play a role in determining galaxy
properties at least since $z=1$. Sample variance due to the finite
survey volume does affect our ability to measure absolute redshift
trends as the number densities vary significantly due to large-scale
structures (particularly evident at $z=0.3-0.4$), but the relative
fractions of different populations should be less affected.

The central panel of Figure~\ref{fig:candy} shows a few examples of
the three morphological classifications from ZEST, chosen randomly to
span the range of colors seen from redshifts $z=0.4-0.6$ and with
$\log(M_{\star}/M_{\odot})>10$. While these classifications do not
always agree with one's visual impression, there are clear differences
in structural parameters among the classes, and correlations between these automated
measurements and a galaxy's position in a group can provide an
interesting test of the dependence of morphology on environment. We
emphasize that the ZEST morphologies are correlated with, but not
identical to, traditional visual classification of ellipticals and
spirals. We compare results from multiple morphological indicators in
Section~\ref{s:systematics}.

\subsection{Radial Trends: Blue Late Disks into Red Bulge+disks}
\label{s:rtrends}

\begin{figure}[htb]
\epsscale{1.2}
\plotone{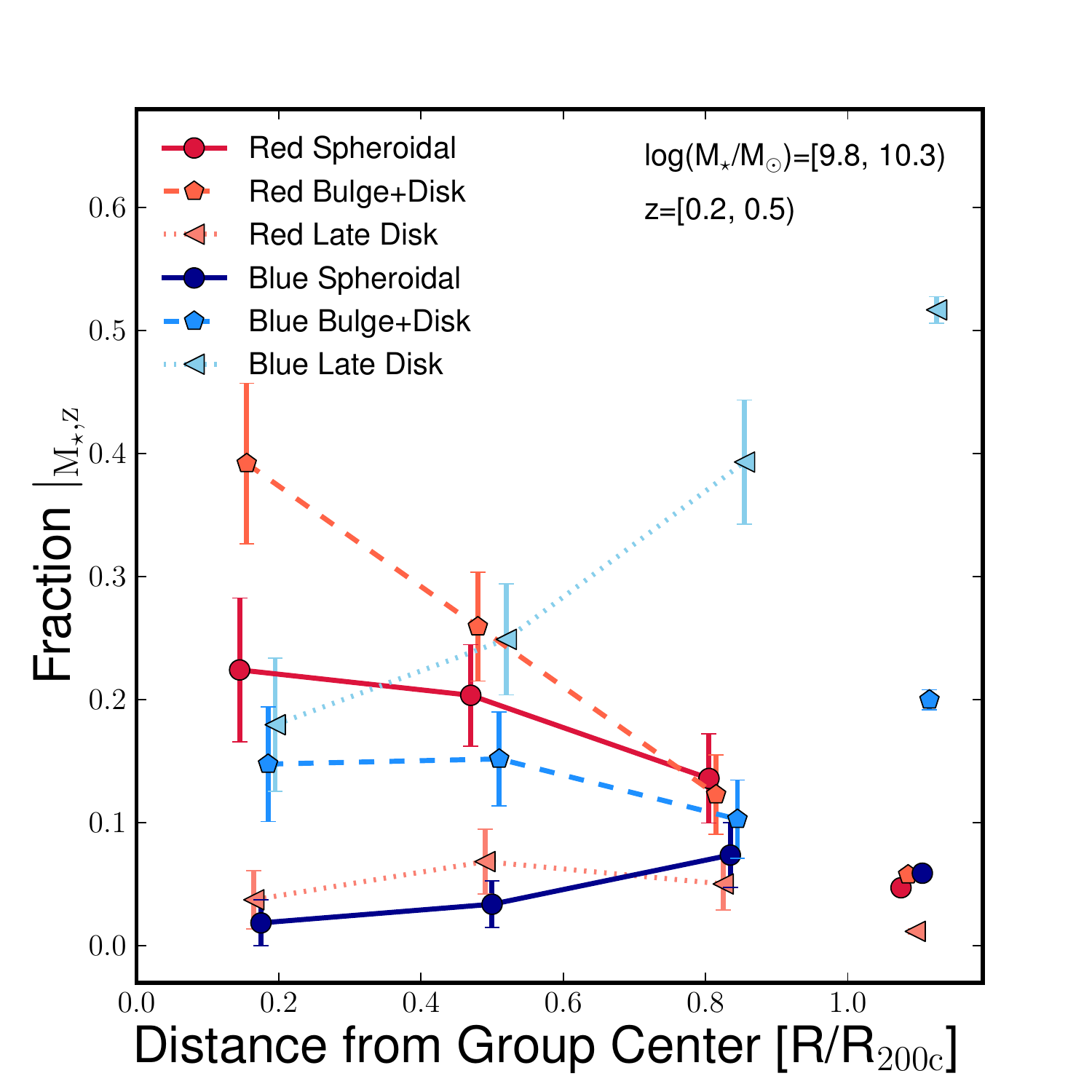}
\caption{Color and morphological fractions as a function of
  group-centric distance. These fractions are calculated for
  satellites in three equally-spaced bins out to $R_{\rm 200c}$ after
  applying contamination corrections. The field population is plotted
  to the right. Points are assigned small horizontal offsets for
  clarity. Error bars are the $1\sigma$ standard deviation of 500
  bootstrap samples. There is a clear transition from blue late disks
  dominating among field galaxies and outer satellites to red
  bulge+disks among inner satellites.}
\label{fig:satrad_single}
\end{figure}

We can quantify some of the trends from Figure~\ref{fig:candy} by
measuring the fraction of galaxies of a given color and morphology as
a function of stellar mass, group-centric distance, and
redshift. Figure~\ref{fig:satrad_single} shows the fraction of
galaxies in each of the six combinations of color and morphology
categories described in Section~\ref{s:otherdata}. For example, the
cyan triangles represent the fraction of galaxies in the stellar mass
and redshift range shown that are both blue and have late disk
morphology. This population makes up the majority among field galaxies
(shown at $R > R_{\rm 200c}$) but its fraction declines among
satellites, contributing less than $20\%$ of the satellite population
at $R < R_{\rm 200c}/3$. Meanwhile, the proportion of red bulge+disk
galaxies rises from $7\%$ in the field to $40\%$ among satellites in
the inner radial bin.

\begin{figure}[htb]
\epsscale{1.25}
\plotone{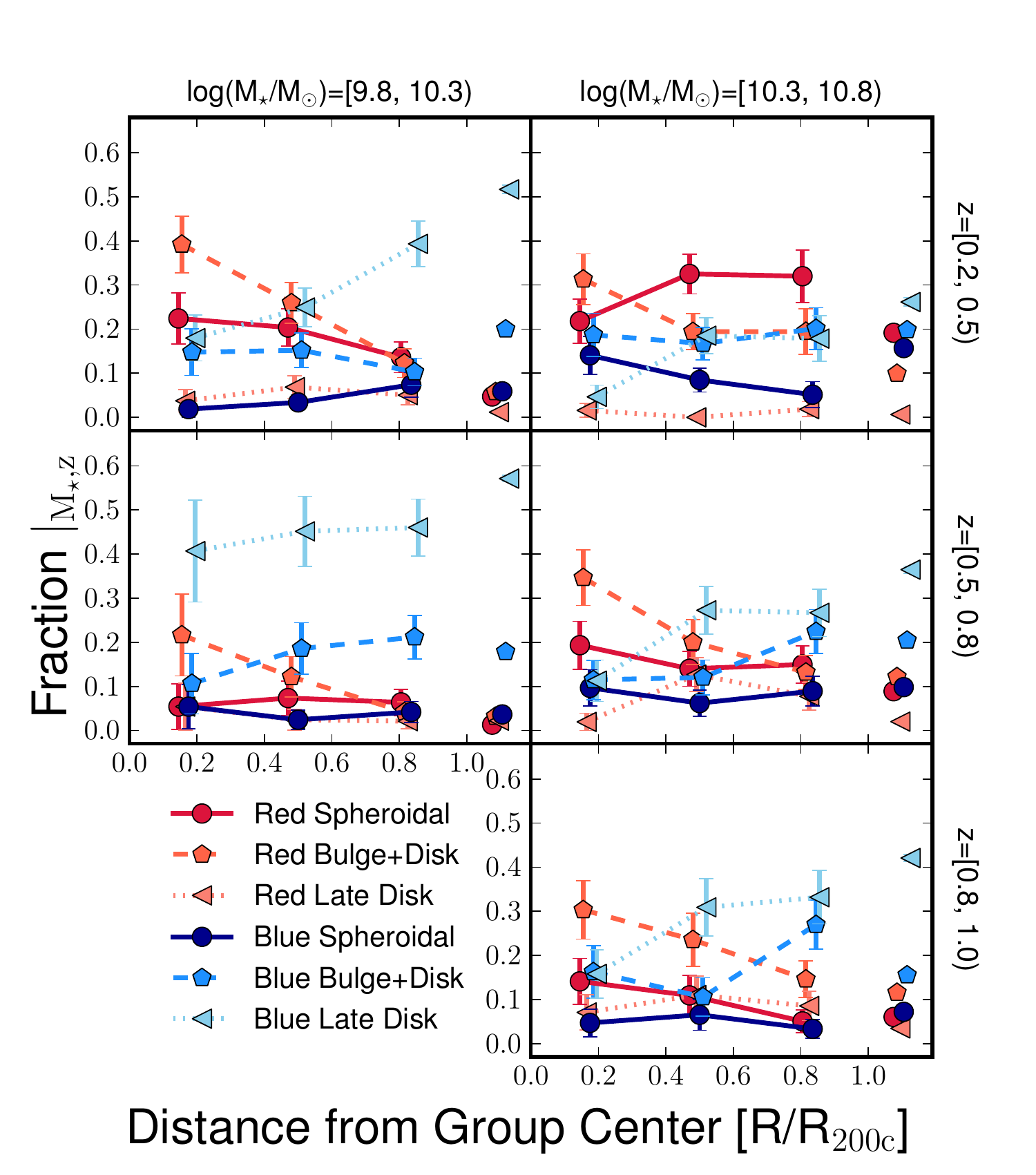}
\caption{Color and morphological fractions of group members as a
  function of distance from the group center, for different bins of
  stellar mass (columns) and redshift (rows). Line styles and error
  bars are as defined in
  Figure~\ref{fig:satrad_single}. Figure~\ref{fig:satrad_single} is
  repeated in the top left panel, for comparison with weakening
  environmental trends at higher mass and redshift.}
\label{fig:satrad}
\end{figure}

Figure~\ref{fig:satrad_single} highlights the most significant
environmental trends in our sample by focusing on low mass, low
redshift galaxies. We repeat the exercise in Figure~\ref{fig:satrad}
with higher mass and redshift bins to study how these trends vary. The
broad picture is similar; blue late disks dominate in the field while
the red bulge+disk population becomes more prominent toward group
centers. Among low mass galaxies at $z>0.5$, the blue late disks
dominate everywhere, while more massive galaxies at lower redshift
have a substantial population of red spheroidals. Red late disks and
blue spheroidals do not make up a large portion of galaxies at
any mass or redshift studied. The red fraction can be determined from these plots by
  summing the three red lines, and similarly the spheroidal fraction
  is the sum of the solid lines with circular markers. The red
  fraction rises toward group centers for low mass galaxies, but is
  relatively flat among massive galaxies. We note that the abscissa for these plots is
the projected group-centric distance, and that the true radial trends measured
in spherical shells are likely more significant than observed.

\subsection{Redshift Trends}
\label{s:ztrends}

\begin{figure}[htb]
\epsscale{1.2}
\plotone{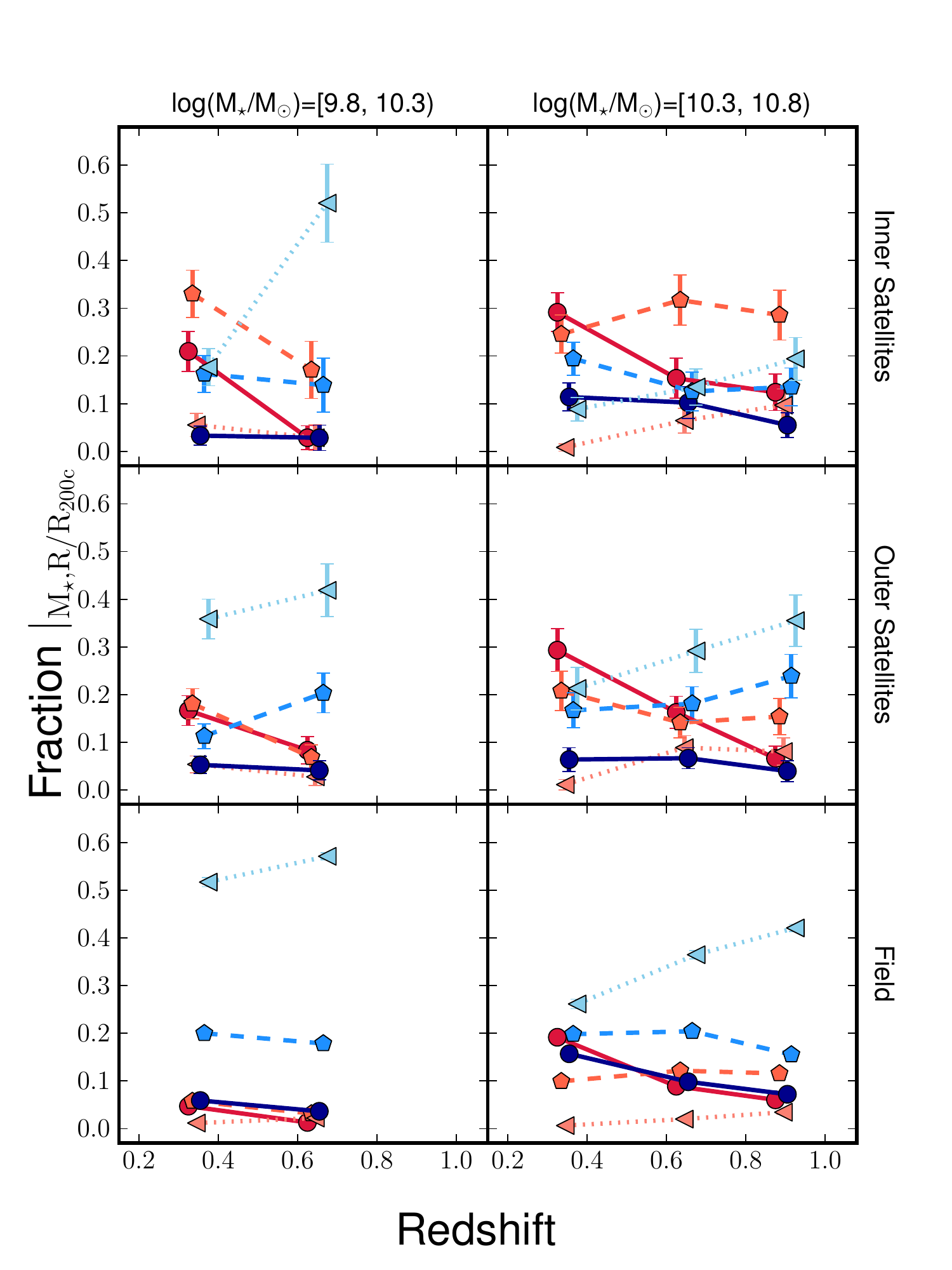}
\caption{Color and morphological fractions as a function of redshift,
  for different bins of stellar mass (columns) and environment
  (rows). Inner and outer satellites are separated at a projected
  group-centric distance of $0.5 R_{\rm 200c}$. Line styles and error
  bars are as defined in Figure~\ref{fig:satrad_single}.}
\label{fig:satz}
\end{figure}

Since satellites tend to fall toward halo centers, group-centric
distance is related to the timescale that galaxies have been inside
the group. The range of redshifts sampled with this data set provides
another measure of time to study evolution. Figure~\ref{fig:satz} is a
transpose of Figure~\ref{fig:satrad} to show the redshift trends in
the distribution of colors and morphologies. Again, there is a decline
in blue late disks among low-mass satellites near the centers of
groups, now compensated by a rise in both red spheroidals and red
bulge+disks. This trend weakens away from group centers and at
higher masses.

\subsection{Checks for Systematics}
\label{s:systematics}

\begin{figure}[htb]
\epsscale{1.25}
\plotone{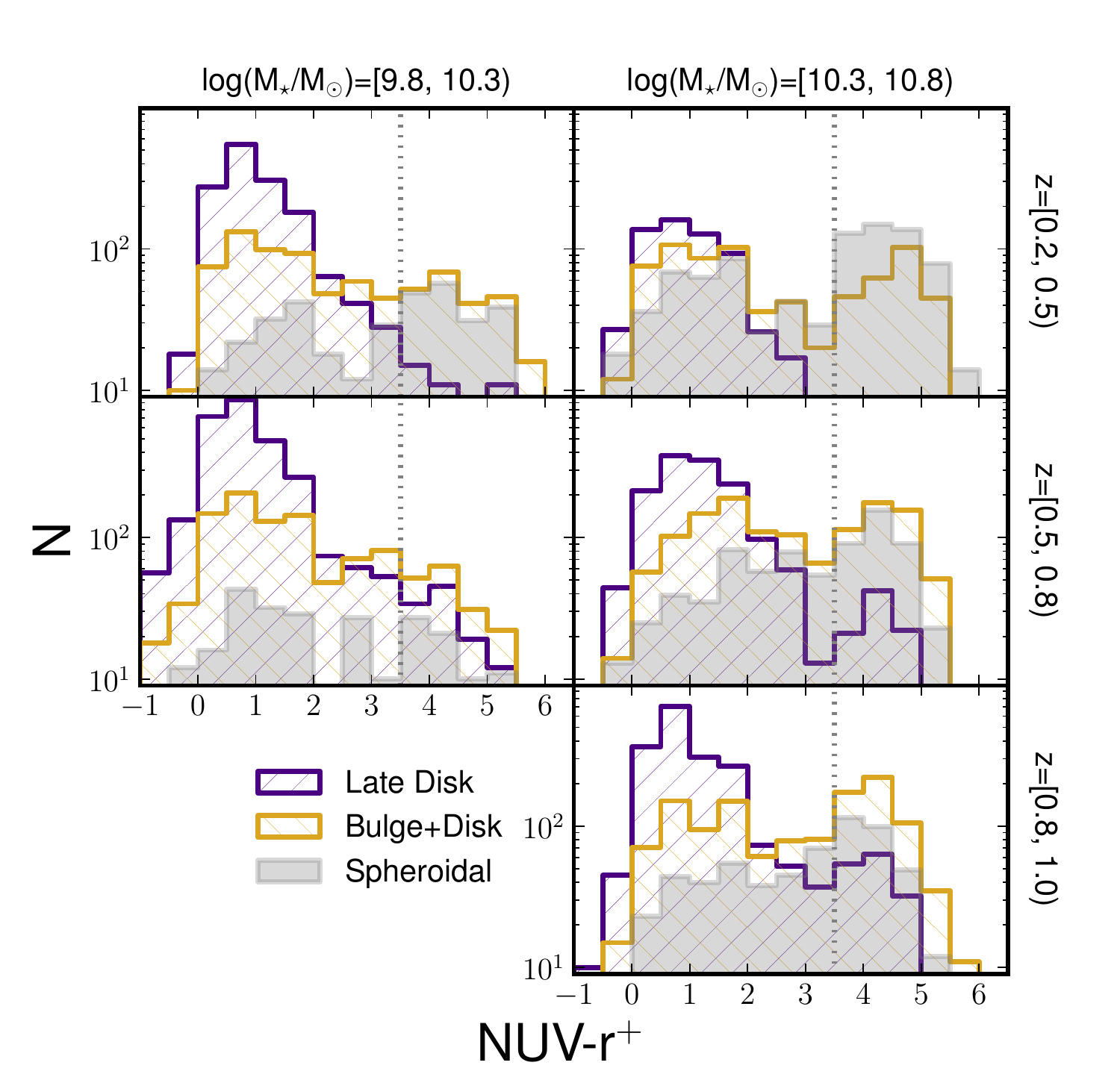}
\caption{Distribution of rest-frame, extinction-corrected, template
  colors by morphological type. Galaxies from all environments are
  included. Vertical dotted lines show the cut used to segregate red
  and blue galaxies.}
\label{fig:color_hist}
\end{figure}
 
\subsubsection{Environment}

There are several possible biases or other effects in the data to
consider in order to ensure the robustness of these results. First, we
revisit the contamination of our satellite sample due to interloping
field galaxies, discussed in Section~\ref{s:groupdata}. In
Figures~\ref{fig:satrad_single},~\ref{fig:satrad}, and~\ref{fig:satz} we
have plotted values of population fractions for satellites corrected for contamination estimated
from mock catalogs. Contamination corrections are always smaller
than the statistical error bars estimated via bootstrapping, except
for low mass blue disk galaxies where it is $10\%$ greater than the
error because the field population is so large. Though the sample of satellites near $R_{\rm 200c}$ is
significantly contaminated by field galaxies, the corrections to the
relative fractions of each galaxy type are small because the field
populations are not markedly different from the outer satellites.

\subsubsection{Color Distribution}

When classifying galaxies by their spectral energy distributions (SEDs), our primary aim
is to distinguish star-forming galaxies from those that are
quenched. Though traditional indicators from emission lines or
spectral breaks are not directly available from photometry, the 31-band SEDs
used here provide a wealth of information about spectral types,
including an estimate of dust extinction that separates star-forming
galaxies that appear red due to dust from those that are truly
passive. The template-based, extinction-corrected \nuvr colors used in
this paper generally have a bimodal distribution with the color cut
from Section~\ref{s:otherdata} falling on the red end of the ``green
valley.'' The color distributions for different morphological types
are shown in Figure~\ref{fig:color_hist}. Shifting the cut slightly in
either direction shifts the amplitude of the red fraction in
Figures~\ref{fig:satrad_single},~\ref{fig:satrad}, and~\ref{fig:satz}
up or down, but the trends with group-centric distance and redshift do
not vary significantly. We have tested an alternative ``red sequence''
selection suggested by \citet{Ilbert2010}, ${\rm NUV - r^+} >
0.5\log(M_{\star}/M_{\odot}) - 0.8 z - 0.5$, using rest-frame absolute
magnitudes but applying no extinction correction. The transformation
of blue late disks into red bulge+disks among low mass satellites is
still evident. Similarly, applying the two-color cut in the ${\rm NUV
  - r^+}$, ${\rm r^+ - J}$ plane used by \citet{Bundy2010} does not
qualitatively change our results.

\subsubsection{Morphologies}
\label{s:sys_morph}

Morphological classification is a challenging problem and significant
scatter exists between the types assigned to galaxies in both visual
and automated analyses. In general, visual analyses emphasize the
presence or absence of spiral features, categorizing objects as
spirals or ellipticals, often with an intermediate class of S0s
grouped with ellipticals. Automated analyses measure structural
parameters such as concentration and asymmetry which are then
generally tied to a training set of visual classifications. The
correlation between the properties measured is imperfect, so we test
the impact on our results of using a variety of automated
morphological classifications, all based on the ACS F814W imaging. The
alternative classifications come from \citet{Tasca2009}, which
presented three separate techniques.

The differences between the results of each catalog and those from the
ZEST classification used in this paper are driven by how bulge+disk
galaxies are classified, since the other catalogs do not split the
spiral/disk category into multiple bins as ZEST does. For instance,
among the $237$ satellites used for Figure~\ref{fig:satrad_single},
the ZEST classifications are $22\%$ spheroidal, $36\%$ bulge+disk,
$37\%$ late disk, and $5\%$ irregular or unclassified. The three separate
classifications for these galaxies from \citet{Tasca2009} vary between
$39 - 59\%$ E/S0s and $59 - 40\%$ spirals.
When bulge+disk galaxies are mostly classified as E/S0s, the
spheroidal populations (both blue and red) in Figures~\ref{fig:satrad}
and~\ref{fig:satz} are significantly elevated but show similar trends
with group-centric distance and redshift. In the opposite case where
bulge+disk galaxies are mostly treated as disks, the spheroidal
fraction is essentially unchanged. While the dominant population in
each panel of Figure~\ref{fig:satrad} and~\ref{fig:satz} can change
depending on which category the intermediate bulge+disk galaxies fall
into, the radial trend in low mass satellites is unchanged; the blue
late type fraction declines toward group centers and is compensated
with a rise in red early types. The redshift trend for this transition
is strongest for low mass satellites near group centers and weakens
toward larger radii and stellar masses. Though the scatter between
morphological classifications signals that our results should be
interpreted and compared to others with caution, the significant
trends with group-centric distance and redshift for the intermediate
bulge+disk galaxies suggests that the ZEST classification has
identified a population in transition.

\subsubsection{A Population of Blue Spheroidals}

While the tests described above suggest that our measures of
environment, color, and morphology are robust, we do note a puzzling
population of massive blue spheroidals that can be seen most clearly
in the bottom right panel of Figure~\ref{fig:satz} as well as the
right column of Figure~\ref{fig:color_hist}. Those plots suggests
that half of all spheroidals in that stellar mass range are blue,
exceeding measurements in other studies \citep[e.g.][]{Kaviraj2007,
  Kaviraj2008, Bamford2009, Schawinksi2009, Ilbert2010}, but see also
\citet{Cross2004} who found a large fraction of blue ellipticals in a
luminosity-selected sample at moderate redshift. We see a similarly
large blue fraction among spheroidals at higher masses
($\log(M_{\star}/M_{\odot}) > 10.8$, not plotted) where spheroidals make up a
higher proportion of all galaxies, and most prominently at low
redshift ($z<0.5$). Visual inspection of the ACS images of these
galaxies suggest that $30-40\%$ show spiral or irregular features,
although the structural parameters measured by the automated
classifiers are all consistent with the red spheroidal population. We
have also investigated the UV-optical and optical-IR colors of these
massive blue spheroidals prior to template fitting, in addition to
optical spectra for $30$ of these objects described in
\citet{George2011}. These data suggest that blue spheroidals are not
typical star-forming galaxies but may have a small amount of recent
star formation to which rest-frame UV measurements are particularly
sensitive \citep[e.g.][]{Kaviraj2007, Kaviraj2008}. Another possible
explanation is that there are unusual stellar populations that are not
represented in the template fits used to derive our \nuvr colors
\citep[see e.g.][for a discussion of galaxy properties contributing to
the ``UV upturn'']{Smith2012}. While this population is interesting in
its own right, it is most prominent at high stellar masses and does
not show a strong dependence on environment, so we do not consider it
further here.

\section{Discussion and Conclusions}
\label{s:discussion}

Our results indicate a complex relationship between color, morphology,
stellar mass, and group-centric distance. Trends in color and
morphology are distinct, and the use of a single property as proxy for
galaxy type does not capture the whole picture. The most interesting
trend seen is the shift in dominance among the low mass population
from blue late disk galaxies in the outskirts to red bulge+disk types
in group interiors. If part of one population was merely disappearing
from the sample, either due to changes in mass or merging with other
galaxies, then the other populations would all be expected to grow by
an equal factor. The fact that the decline in one population is
roughly balanced by the rise of a single other population suggests a
transformation process. We now discuss these results in the context of
past analyses, the physical implications of the present work, and
future avenues to clarify the role of environment in galaxy evolution.

\subsection{Connection to Previous Observations}

There is a long history of research into the covariance between the
stellar masses, colors, morphologies, and environments of galaxies and
its evolution with time. After controlling for differences in stellar
mass or luminosity, numerous studies at low redshift have found that color is more
strongly correlated with environment than morphology
\citep[e.g.,][]{Kauffmann2004, Blanton2005, Christlein2005, vandenBosch2008,
  Bamford2009, Skibba2009, Weinmann2009}. The implication of these
studies is that the well-known correlation between morphology and
environment is secondary to correlations between morphology and
stellar mass or color, with the latter properties more physically
linked to environment. Still, some of these studies find residual
correlations between morphology and environment after controlling for
color and stellar mass or luminosity, particularly at low masses and
among late to intermediate morphologies \citep{Blanton2005,
  Weinmann2009, Skibba2012}.

Moving from large low-redshift studies to moderate redshifts ($z\sim0.4$), \citet{Balogh2009} and
\citet{McGee2011} have shown that groups still host a higher fraction of red
galaxies than the field, while \citet{Wilman2009} found
an elevated S0 fraction in groups with a mildly significant rise with
group-centric distance. This radial trend in morphology is the opposite of that seen
in Figure~\ref{fig:satrad} but may be due to a smaller sample size or
the use of luminosity-weighted centroids, which have been shown to be
a poorer tracer of halo centers measured by weak lensing when compared
with the most massive galaxy used in this catalog \citep{George2012}.

Environmental trends at fixed stellar mass have been seen to weaken at
higher redshift \citep[e.g.,][]{Poggianti2008, Tasca2009,
  Cucciati2010, Iovino2010, Kovac2010}, though recent analyses have
shown clear differences between the fraction of star-forming galaxies
across environments over a range of stellar masses at least up to $z\sim1$
\citep{Cooper2010, Peng2010, George2011, Knobel2012}. \citet{Tasca2009} and
\citet{Kovac2010} have compared the dependence of colors and
morphologies on local density and in groups in the COSMOS field,
finding a stronger effect on color than on morphology, similar to the
low redshift results and suggesting a longer timescale for structural
transformations than for quenching star-formation. \citet{Muzzin2012}
and \citet{Presotto2012} have also shown that the quenched fraction
among satellites depends significantly on group-centric distance at
these redshifts.


The results presented here extend this work to include radial trends
in colors and morphologies from $z=0.2-1$. Our X-ray group catalog
gives a large, clean, and fairly representative selection of
$10^{13}-10^{14}~{\rm M_{\odot}}$ halos \citep{Finoguenov2010}, and
has been well-calibrated based on 
its weak lensing signal \citep{Leauthaud2010, George2012}. We detect significant
trends in both color and morphology with group-centric distance. Some previous studies measured weak or insignificant
gradients in morphology by using a simple dichotomy of spirals and
ellipticals, and we can reproduce these results when using a coarse
morphological binning (see discussion in Section~\ref{s:sys_morph}). But in contrast
to those results, we see clear morphological gradients at
fixed stellar mass and color once the morphological classification
considers differences in the bulge content of disk galaxies.

The importance of this intermediate morphology between pure disks and
spheroidals has been noted before in the context of S0 galaxies which
have been known to dominate the evolution in the morphology-density
relation \citep{Dressler1997, Postman2005, Smith2005, Boselli2006b,
  Moran2007, Oesch2010, Lackner2013}.
\citet{Smith2005} also showed that morphological evolution occurs
later in less dense regions than in the densest regions associated
with clusters, with little evolution at intermediate densities from
$z=1$ to $0.5$ followed by an increase in early types at lower
redshift. There is a significant overlap between the intermediate
bulge+disk population in our analysis and visually classified S0
galaxies although these bulge+disk galaxies are often classified
slightly later along the Hubble sequence \citep{Scarlata2007}. Our
results in Figure~\ref{fig:satz} broadly corroborate those of
\citet{Smith2005} and add that color and stellar mass are also
important dimensions when studying these trends.


\subsection{Implications for Physical Mechanisms of Galaxy Transformation}

The radial gradients and redshift trends measured in
Section~\ref{s:results} suggest a transformation among low mass
satellites from blue late disks into red bulge+disks and spheroidals.
The mechanism for this transformation must affect both color and
morphology, or more physically, the star formation rate and stellar
kinematics. Some processes \citep[see e.g.][for a review]{Boselli2006}
halt star formation without significantly altering stellar structure,
such as gas removal via ram pressure stripping or weak tidal
interactions, suppression of gas accretion within dense
shock-heated environments, or quasar feedback. On the other hand, galaxy mergers and
strong tidal interactions can affect both the distribution of gas
needed to form stars as well as the stellar morphology. A third
possibility is that color and morphology changes are physically
coupled; bulge growth could stabilize a galaxy against
  disk fragmentation and suppress star formation \citep{Martig2009},
  or gas loss may leave a disk unable to dissipate energy from tidal
  interactions or may drive instabilities leading to bulge growth.
Yet another scenario is that a bulge only appears more prominent after
a galaxy is quenched because the previously star-forming disk has
faded.

\begin{figure}[htb]
\epsscale{1.2}
\plotone{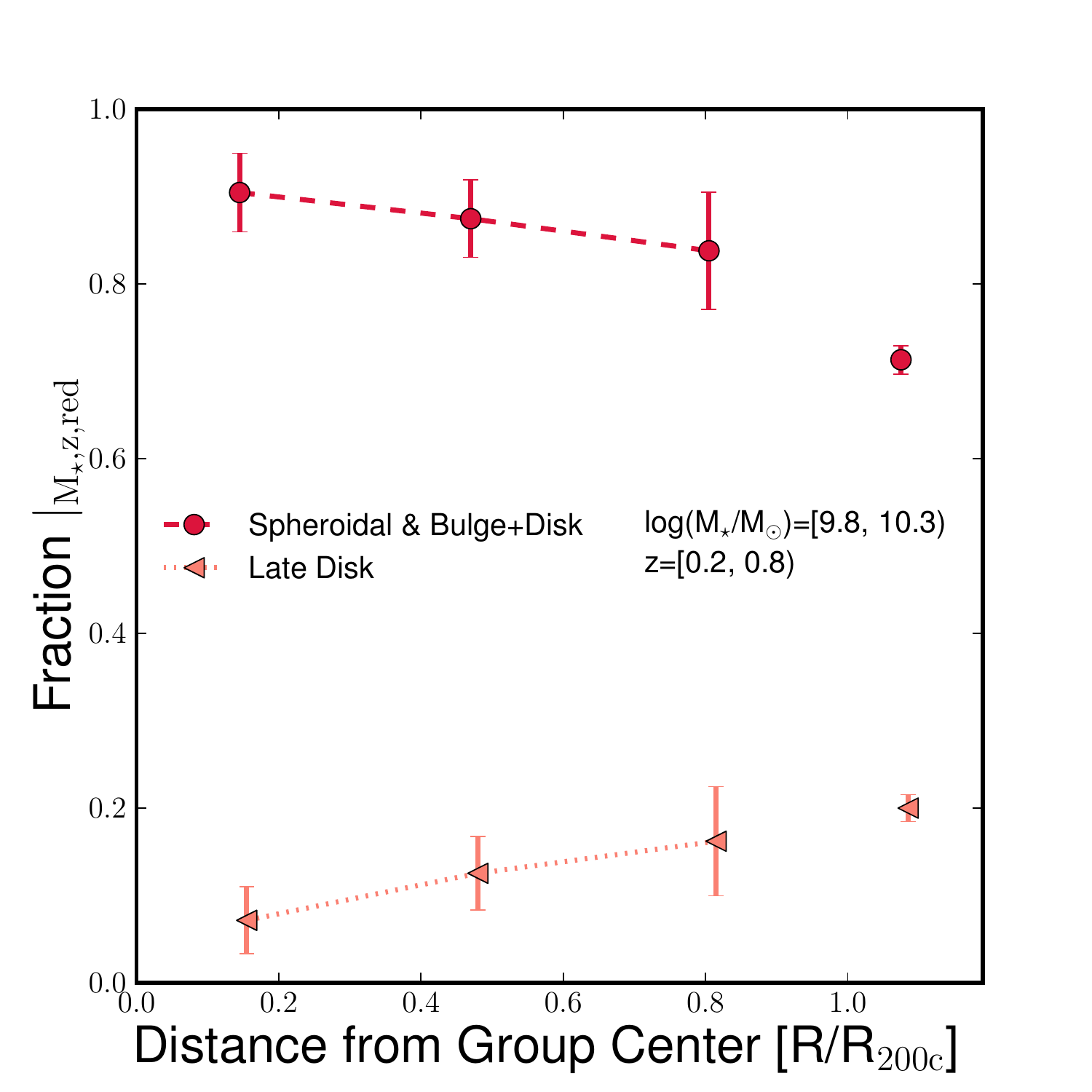}
\caption{Morphological fractions \textit{among red galaxies} as a
  function of distance from the group center. The small but
  significant excess in early-type morphologies among inner satellites
  relative to field galaxies suggests that mergers or tidal
  interactions cause more significant bulge growth among
  satellites.}
\label{fig:satradred}
\end{figure}

We can test these models by studying the morphological dependence on
environment among quenched galaxies. Mechanisms like gas stripping or
disk fading that do not directly affect morphology should produce a
higher fraction of quenched galaxies in dense environments, but among
quenched galaxies the morphological distribution should be
constant across environments. On the other hand, if quenched galaxies in dense
environments have more bulge-dominated morphologies than quenched
field galaxies, it would suggest that mergers or strong tidal
interactions altering the structure of galaxies occur in addition to,
or in conjunction with, the suppression of star formation. The results
of this test are plotted in Figure~\ref{fig:satradred}, where we show
the fractions of red galaxies that are early and late type as a
function of environment. We have used a broad redshift range and
combined the spheroidal and bulge+disk categories to reduce
statistical errors for this smaller population of red
galaxies. Figure~\ref{fig:satradred} demonstrates that quenched
satellites, particularly those in the inner regions of groups, are
more likely to be bulge-dominated than their field
counterparts. 

We interpret this to mean that some physical process in dense
environments is driving bulge growth. Though some field studies \citep{Bundy2010, Masters2010}
have already noted a tendency of passive disks to be more concentrated than
star-forming disks, we find that the morphological evolution of
quenched galaxies is even stronger in groups. Our result is consistent with previous
arguments favoring bulge growth over disk fading based on the higher
typical luminosities of galaxies with intermediate morphologies
compared to late types \citep{Christlein2004,Burstein2005}. The data
presented here show that even at fixed mass there must be bulge enhancement
that accompanies quenching in groups.

\begin{figure*}[htb]
\epsscale{1.1}
\plottwo{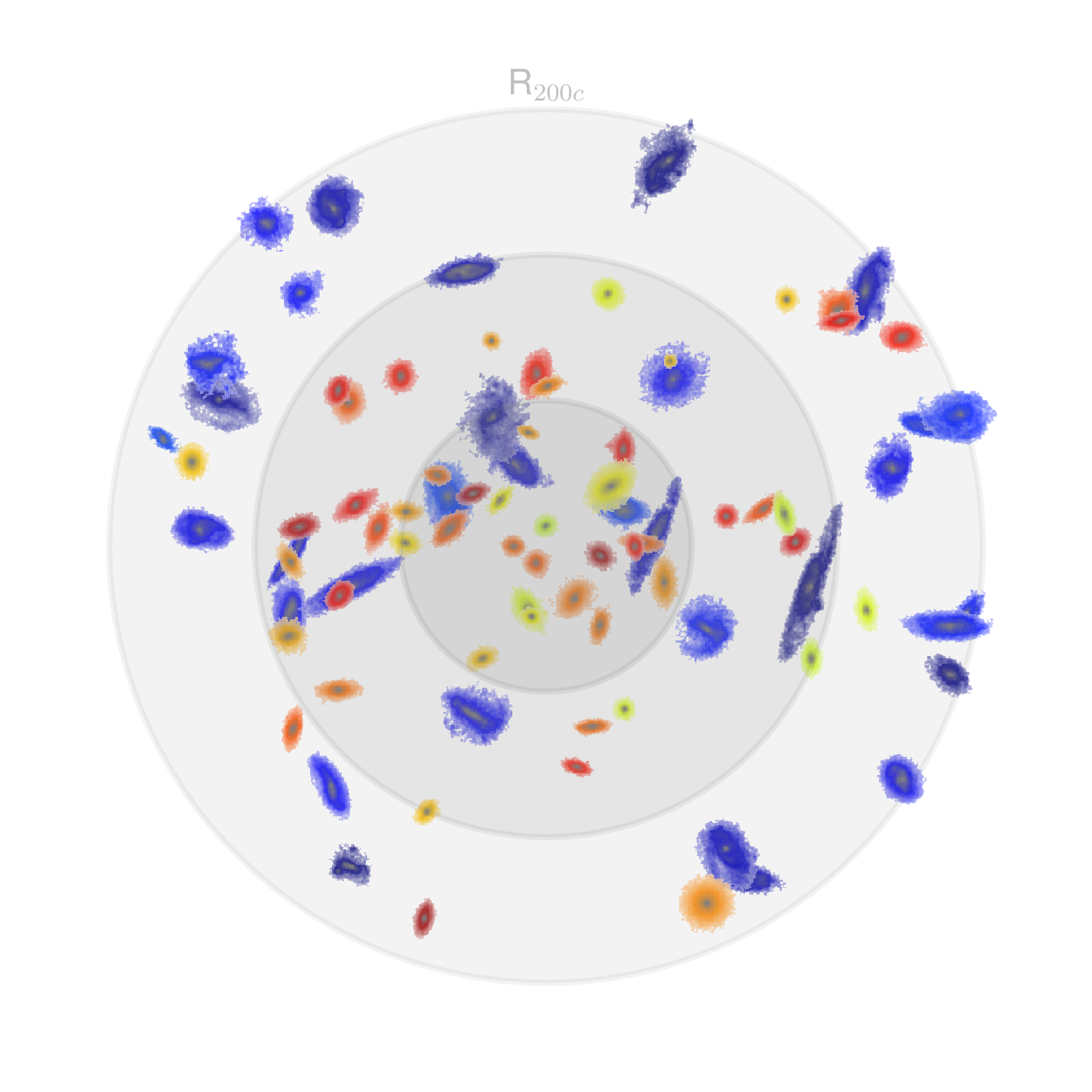}{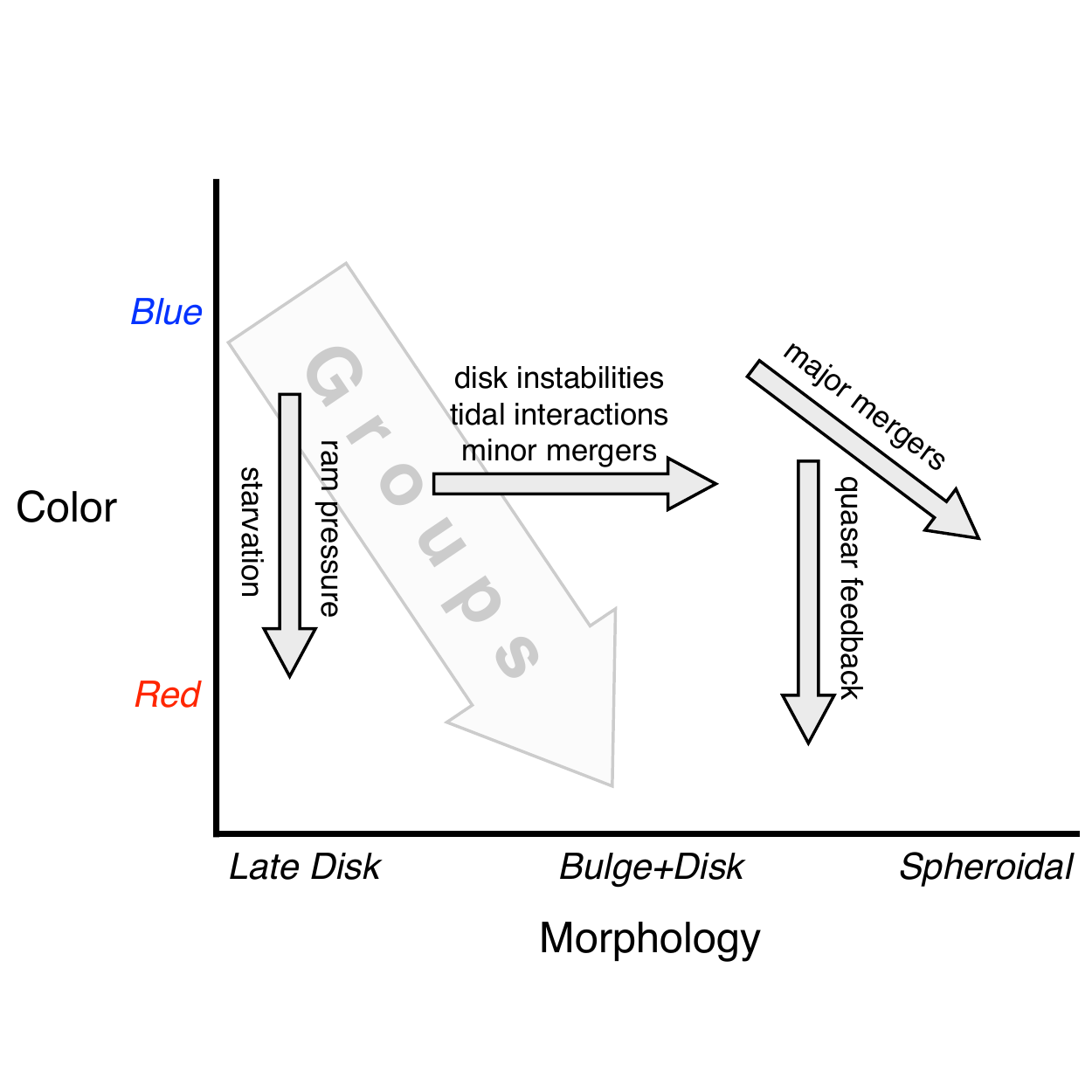}
\caption{Illustrations of our main results and interpretation. Left
  panel shows the projected positions of satellites in an ensemble
  group with the same stellar mass and redshift range as
  Figure~\ref{fig:satrad_single} with only the blue late disks (which
  dominate the outskirts) and red bulge+disks (which dominate the
  interior) displayed. The schematic diagram at right shows the effects of various
  physical mechanisms on color and morphology; the large arrow
  indicates the observed transformation from blue late disks to red
  bulge+disks, suggesting a combination of processes.}
\label{fig:cartoons}
\end{figure*}

The results of Figure~\ref{fig:satrad} suggest that the timescale for
the morphological transition from late disk to bulge+disk must be
comparable to the timescale for quenching in order to turn blue late
disks directly into red bulge+disk galaxies. For example, if quenching
occurred via gas stripping or removal at a rate faster than any
structural changes, we would expect to see blue late disks turning
into red late disks. Instead we observe a growth in red galaxies with
earlier morphologies whenever blue late disks decline.  The fraction
of blue late disks plummets by more than a factor of two in
$\sim2~{\rm Gyr}$ among inner satellites (top left panel of
Figure~\ref{fig:satz}), while the fraction of red late disks is nearly
constant and the fraction of red bulge+disks grows. Similarly, we do
not detect a large population of blue bulge+disk galaxies which might
be expected if bulge growth happened faster than quenching. These
observations suggest that some morphological evolution occurs on a
similar timescale as quenching.

At the same time, the morphological transformation observed at low
stellar masses has not fully turned many disks into spheroidals. The
bulge+disk galaxies are still categorized as disks in the top level
ZEST classification.  This echoes earlier results finding a buildup of
S0s in clusters but weaker evolution in the relative abundance of
ellipticals \citep[e.g.,][]{Dressler1997}. Though the fraction of red
spheroidals does not correlate significantly with group-centric
distance (Figure~\ref{fig:satrad}), it does grow globally with time
(Figure~\ref{fig:satz}). This may indicate that different processes
are responsible for producing bulge+disks and spheroidals. 

The discreteness of morphological classification makes detailed
comparison of transformations in morphology and color somewhat
difficult, but the structural evolution appears to contrast with that
in color, where a clear bimodality separates red and blue states and
the relatively low intermediate population suggests a fast quenching
timescale once it begins (e.g., \citealt{Wetzel2012a}, though see
\citealt{Balogh2011}). One could test the hypothesis that galaxies pass through an
intermediate state between blue late disk and red bulge+disk by
measuring the time since quenching based on the SEDs of galaxies in
these intermediate states. Is there a difference in the mean color of
blue bulge+disks and blue late disks, or between red bulge+disks and
red late disks? Figure~\ref{fig:color_hist} does not show conclusive
differences in the colors of these populations, and the satellite
sample size is too small to constrain differences in the distribution
of colors within the red and blue populations. More detailed analysis
with spectroscopic data could better constrain such evolutionary
models.

The group environment can play an important role in building up the
well-studied color-morphology-density relations in nearby clusters. A
group of mass $10^{13.5} M_{\odot}$ at $z=1$ should grow through
mergers by an average of $0.3$ dex to $z=0$ \citep{Fakhouri2010}, so
some of our high redshift groups will be the progenitors of massive
clusters and others may accrete onto them. For comparison, a typical
star-forming galaxy should grow by $0.2$ dex in stellar mass between
each of our three redshift bins and by $0.7$ dex from $z=1$ to $0$
\citep{Elbaz2011}.  \citet{Wetzel2012b} show that the dominant
population of satellites in the most massive halos at $z=0$ were
already satellites in groups when accreted. Our results demonstrate
that both the color and morphology of these satellites are influenced
in the group environment, showing the importance of ``preprocessing''
of galaxies prior to entering massive clusters. The bulge+disk
population we see here is not quite as morphologically evolved as the
growing S0 population seen at low redshift and in massive clusters,
but likely precedes it.
 
We summarize our main results and interpretation in
  Figure~\ref{fig:cartoons}. The left panel shows the positions of
  satellites for a stacked group and highlights the trend in
  Figure~\ref{fig:satrad_single} with blue late disks dominating the
  outskirts and red bulge+disk galaxies making up most of the inner
  satellite population. This combination of color and morphology
  transformations suggests some combination of gas removal and bulge
  growth, shown in the right panel. While the diagram is a
  simplification of a wide variety of model predictions, the primary
  point is that \textit{both} color and morphology are affected in
  group satellites, so physical mechanisms that explain environmental
  correlations should reflect this.




\subsection{Future Prospects}

While these measurements add insight to the transformation of galaxy
properties, much work remains to be done to constrain the variety of
physical mechanisms that may be responsible. A cleaner bulge-disk
decomposition would be useful to track the growth of bulges and the
importance of disk fading \citep[e.g.,][]{Lackner2012,
  Lackner2013}. Incorporating the physical sizes of each component
would also help to link these transformations with the significant
growth seen in early types since $z\sim2$
\citep[e.g.,][]{Bruce2012}. With deep high-resolution imaging from
\textit{HST}, the CANDELS survey \citep{Grogin2011,Koekemoer2011} is pushing these
studies to higher redshift and should also allow bulge-disk
decomposition to be studied at multiple wavelengths, identifying when
and where star formation is happening or stopping. Alternatively,
visual classifications from an ongoing Galaxy Zoo project in the
CANDELS fields will track the evolution of spiral features and bars
which are challenging for automated techniques. Impending wide field
imaging surveys will add greatly to the statistics of these analyses.

Different measures of environment may also help disentangle the
relevant physical mechanisms. While we have argued in favor of
halo-based indicators (see \citealt{George2011} for a discussion), the
local galaxy density \textit{within a halo}, in addition to
group-centric distance, may shed light on galaxy-galaxy interactions
and infalling substructure \citep[e.g.,][]{Blanton2007, Cibinel2012b,
  Woo2012}. However, this indicator is a noisy quantity affected by
shot noise, redshift errors, and peculiar velocities, likely
requiring deep and complete spectroscopic data for clean
results. Larger surveys will also enable a wider range of halo masses
to be probed at higher redshift. Comparison of different halo mass
proxies can provide a test of environmental mechanisms, for example if
X-ray bright groups are more efficient at ram pressure stripping
satellites than groups with less hot gas.

These studies can also be extended to the interplay of environmental
mechanisms with non-stellar components of galaxies, namely the gas
content and central black holes in galaxies. Current and planned radio
arrays will extend the study of neutral and molecular gas beyond the
local Universe allowing a clearer picture of how star-formation is fed and
quenched. And the growth of bulges seen in this paper should be
connected with accretion onto the central black hole in order to
explain the tight correlation seen locally between these components.

Finally, while we have simply presented an empirical description of
the data here, we can extend this work by modeling the evolution of
color and morphology with environment. \citet{Peng2010} and
\citet{Wetzel2012b} present simple empirical descriptions of the
fraction of quenched galaxies as a function of stellar mass and
environment. Incorporating morphologies into this framework will
further illuminate the physical processes at work that build up the
environmental correlations we observe.


\acknowledgments 

We thank Andrew Wetzel, Frank van den Bosch, and Nic Ross for helpful
discussions, and Claire Lackner for constructive comments on a draft.
MRG acknowledges support from the US Department of Energy's Office of
High Energy Physics (DE-AC02-05CH11231) and a Graduate Research
Fellowship from the US National Science Foundation. CPM is partially supported by a grant from
the Simons Foundation (\#224959). This work was also
supported by World Premier International Research Center Initiative
(WPI Initiative), MEXT, Japan. We gratefully acknowledge the
contributions of the entire COSMOS collaboration. More information on
the COSMOS survey is available at \url{http://cosmos.astro.caltech.edu}.  This research has made use
of the NASA/IPAC Infrared Science Archive, which is operated by the
Jet Propulsion Laboratory, California Institute of Technology, under
contract with the National Aeronautics and Space Administration.

\mbox{~} 


\bibliographystyle{apj}

\bibliography{transformers}


\end{document}